\documentclass[showpacs,preprintnumbers,amsmath,amssymb,floatfix]{revtex4}
\usepackage{amsmath,graphicx,amsfonts}
\usepackage{dcolumn}

\newcommand{\be}{\begin{equation}}
\newcommand{\ee}{\end{equation}}
\newcommand{\ket}[1]{| #1 \rangle}

\newcommand{\moy}[3]{\langle #1 | #2 | #3 \rangle}

\begin{document}
{
\title{\bf Quantum oscillations in a two-mode atom-molecule
    Bose-Einstein condensate -- the discrete WKB approach} 

  \author{J.  Dorignac} \affiliation{College of engineering, Boston University,
44 Cummington street, 02215, Boston, Massachussets, USA}

  \author{Yu. B. Gaididei} \affiliation{Bogolyubov Institute for
    Theoretical Physics, Metrologichna str. 14 B, 01413, Kiev,
    Ukraine}

  \author{J. C. Eilbeck} \affiliation{Department of Mathematics
    Heriot-Watt University Riccarton, Edinburgh EH14 4A} 

  \author{ P.  L. Christiansen} \affiliation{Informatics and
    Mathematical Modelling and Department of Physics, The Technical
    University of Denmark, DK-2800 Lyngby, Denmark}

  \author{C. Katerji } \affiliation{Department of Applied Physics I,
    University of Sevilla, Avenida Reina Mercedes str, 41012-Sevilla,
    Spain}

\date{\today}
\begin{abstract}
  Quantum effects in a system of coupled atomic and molecular
  Bose-Einstein condensates in the framework of a two-mode model are
  studied numerically and analytically, using the discrete WKB
  approach. In contrast to the mean-field approximation, the WKB
  analytical results are in a very good agreement with numerical
  results. The quantum fluctuations of the atomic and molecular
  populations are calculated, and found to be of the same order of
  magnitude as their mean values.
\end{abstract}
\maketitle

\section{Introduction}

There is a growing interest in super-chemical \cite{heinz} properties
of weakly interacting gases in the Bose-Einstein condensate (BEC)
state.  One of the major goals in this field is to produce a {\em
  molecular} Bose-Einstein condensate.  To date, this has been
achieved by creating it from an {\em atomic} BEC.  Two routes are
mainly used: ultra-cold molecules are formed by photo-association of
atoms in a BEC \cite{wynar,mckenzie} or they are produced by applying
a time-varying magnetic field near a Feshbach resonance
\cite{stenger,cornish,donley}.

One of the first theoretical attempts to describe the production of a
molecular BEC via Raman photo-association of an atomic BEC has been
put forward in Ref.\ \cite{drummond}. There, the possibility of
coherent formation of a molecular BEC is treated using a parametric
field theory.  In Ref.\ \cite{java}, a two-mode quantum Hamiltonian
was proposed, that accounts for the statistics of both the atoms and
the molecules. In this model, large amplitude nonlinear oscillations
between atomic and molecular condensates are observed, that are damped
away at long times. Using the same two-mode model, the dynamical
evolution of an atomic-molecular BEC has been investigated in Ref.\
\cite{vardi} and compared to the mean-field theory predictions. One of
the conclusions drawn in this paper is that, large amplitude
atomic-molecular oscillations, which were predicted in
Ref.~\cite{java} on the basis of mean-field results, are damped by the
rapid growth of fluctuations near the dynamically unstable molecular
mode.  Within the same framework, the bi-stability and quantum
fluctuations in coherent photo-association of a BEC have been
considered in Ref.\ \cite{wu}.  The number statistics of the two-mode
model have also been compared to the number statistics of normal atomic
Fermi gases and of Fermi systems with pair (BCS) correlations
\cite{Meiser05}.  Quite recently a coherent superposition of atoms and
molecules condensate was observed in a Rb Bose-Einstein condensate by
using a stimulated Raman adiabatic passage \cite{winkler}

In this paper, we show that, in the two-mode model proposed in
Ref.~\cite{java}, the expectation value of the population imbalance
between atoms and molecules, as well as the level fluctuations, {\em
  depend drastically on the initial value of the imbalance}.  We study
this problem numerically and analytically.  Our analytical results,
based on the discrete WKB method developed in Ref.~\cite{braun}, are
confirmed by the numerical solution of the exact quantum equations.

\section{The Model}

We consider the simplest model of atomic-molecular condensate,
consisting of an atomic mode $A$ and a molecular mode $B$. Let $b
\,(b^\dagger)$ and $a \,(a^\dagger)$ be the annihilation (creation)
operators for the molecular field $B$ and the atomic field $A$, respectively.
The two modes are coupled coherently by an association-dissociation
process of a Fermi-resonance \cite{herzberg} type, and the effective
Hamiltonian has the form
\begin{eqnarray}\label{hamilt}
  H=\frac{\Delta}{2}\,a^\dagger\,a+{\chi\over 2\sqrt{V}}\,
  \left(b^\dagger\,a a+a^\dagger\,a^\dagger\,b\right)
\end{eqnarray}
where the detuning parameter $\Delta=2\mu_a-\mu_m$ characterizes the
difference between the chemical potentials of the molecular ($\mu_m$)
and the atomic ($\mu_a$) modes. In the case of coherent
photo-association of a Bose-Einstein condensate, the Fermi-coupling
parameter $\chi/\sqrt{V}$ (Rabi frequency) depends on the volume $V$
of the system \cite{kostrun}.

Using Hamiltonian (\ref{hamilt}), the Heisenberg equations for the
operators \footnote{Explicitly time-dependent operators are 
in the Heisenberg representation, $\hat{A}(t)=e^{i\hat{H} t}\hat{A}
e^{-i\hat{H} t}$.}
\begin{equation}
\hat{n}(t)=a^\dagger(t)\,a(t)\, ,\quad \hat{x}(t)=b^\dagger(t)\,a(t)^2 
+a^\dagger(t)^2\,b(t)\quad {\rm and}\quad 
\hat{y}(t)=i\left(b^\dagger(t)\,a(t)^2 -
a^\dagger(t)^2\,b(t)\right),
\end{equation} 
are given by
\begin{eqnarray}\label{heq}
{d\hat{n}(t)\over d t}={\chi\over \sqrt{V}}\hat{y}(t),\quad
{d\hat{x}(t)\over d t}=-\Delta\hat{y}(t),\quad
{d\hat{y}(t)\over dt}=\Delta\hat{x}(t)-{\chi\over \sqrt{V}}\,
\left(3\hat{n}(t)^2-2N\hat{n}(t)-N\right).
\end{eqnarray}
The total number of atoms in the system,
$N = a^{\dagger} a + 2  b^{\dagger} b$,
is a conserved quantity.

In the mean-field approximation, $\langle \hat{n}^2\rangle\approx
\langle \hat{n}\rangle^2$, where $\langle
\dots\rangle$ stands for the average over the initial state of the system.
Neglecting the term $\chi N/\sqrt{V}$ which is of order $N$ times less
than the remaining terms in the right-hand side of Eq.\ (\ref{heq}),
the equations of motion for $\left(x(t), y(t), n(t)\right)$ have 
stationary point $\left(0,0,0\right)$, which corresponds to the entire
population being in the molecular mode.  However, for $\Delta^2 <
2\chi^2\,\nu$, where $\nu=N/V$ is the concentration of atoms in the system,
this state is dynamically unstable \cite{vardi}.  
It is one of the goals of this
paper to go beyond the mean-field approximation and to clarify the behaviour
of an atomic-molecular Bose-Einstein condensate in the limit of
strong Fermi-coupling.

\section{Dynamical evolution of the number of atoms}

In what follows, we shall use the basis of states 
\begin{equation} \label{St1}
\ket{n_a,n_b} =
\frac{(a^{\dagger})^{n_a}(b^{\dagger})^{n_b}}{\sqrt{n_a!n_b!}} \ket{0}\, .
\end{equation}   
Because the total number of atoms, $N$, is conserved,
$\ket{n_a,n_b}=\ket{n_a,(N-n_a)/2}\equiv \ket{n_a}$.  Notice that
$n_a$ and $N$ have same parity. For the sake of simplicity, we
assume $N$ even and put $N=2P$, $n_a=2p$ $(p \in
\{0,1,2,\cdots,P\})$. We look for a solution of the
Schr{\"o}dinger equation $H\,\ket{\phi}=E\,\ket{\phi}$
in the form $\ket{\phi} = \sum_{p=0}^{P} C_p \ket{2p}$.
The coefficients $C_p$ satisfy the set of equations
\begin{equation} \label{TTR}
E C_p = \Delta\, p \, C_p + {2\chi\over\sqrt{V}}\,
\left(f(p) C_{p-1} + f(p+1) C_{p+1}\right)\,,\quad p \in
\{0,1,2,\cdots,P\}, 
\end{equation} 
with  
\begin{equation} \label{funcf}
f(p) = \sqrt{p(p-{\textstyle \frac{1}{2}})(P+1-p)}\, .
\end{equation} 
The number of possible energies and eigenstates, 
$\{E^{(\alpha)},\ket{\phi^{(\alpha)}}\}$, is $P+1$. 
In a system containing initially $n_0 = 2p_0$ atoms, the
expectation value of the fraction of atoms, $z(t)=n(t)/N$, 
evolves according to
\begin{eqnarray}\label{nt}
z(t)={1\over N}  \moy{n_0}{e^{i\hat{H} t}
\hat{n} e^{-i\hat{H} t}}{n_0}=\overline{z} + \sum_{\alpha < \alpha'}
I_{\alpha,\alpha'} (n_0) \cos \, \omega_{\alpha,\alpha'} t\ ,   
\end{eqnarray}
where 
\begin{equation} \label{I00}
\overline{z}\equiv \lim_{T \rightarrow \infty} \frac{1}{T} \int_{0}^{T} z(t)\,
dt\, = {2\over N} \,\sum_{\alpha} 
\left(C_{p_0}^{(\alpha)}\right)^2 \sum_{p} p
\left(C_{p}^{(\alpha)}\right)^2   
\end{equation}
represents the constant (dc-) part of $z(t)$
and where
\begin{equation} \label{Inunup}
I_{\alpha,\alpha'}(n_0) = {4\over N}\,  C_{p_0}^{(\alpha')} 
C_{p_0}^{(\alpha)} \sum_{p} p
C_{p}^{(\alpha')} C_{p}^{(\alpha)} 
\end{equation}
are the respective intensities of the $P(P+1)/2$
frequencies $\omega_{\alpha,\alpha'} = E^{(\alpha')}-E^{(\alpha)}$, 
$\alpha' > \alpha$, appearing in the system.

\section{Discrete WKB solution}

To gain some insight into the dynamics of the system, we use the
so-called discrete WKB-approach (see Ref.\ \cite{braun}) to solve 
the three-term recurrence equation \eqref{TTR} 
in the limit where $P\gg 1$. We introduce a new variable
$x=p/P$ and consider the coefficients $C(p)=C(xP)\equiv c(x)$ as
functions of the continuous variable $x$. Now, up to 
order 2 in $\epsilon\equiv 1/P$, Eq.\ (\ref{TTR}) reads
\begin{equation}
\label{wkb}
\left(\lambda- \delta \,x\right)\,c(x)-
\left(\epsilon\,\left(\partial_x
    F\right)\,\sinh\left(\epsilon\,\partial_x \right)+2
  F\,\cosh\left(\epsilon\,\partial_x \right)\right)\,c(x)=0 
\end{equation} 
where $\lambda=E/\left(\sqrt{2\nu}\,\chi
  P\right)$, $\delta=\Delta/\left(\sqrt{2\nu}\,\chi\right)$ and
$F(x)=P^{-3/2}\,f(x P+1/2)\approx x\,\sqrt{1-x}$.
We now use the WKB method to treat equation \eqref{wkb}. We look 
for a solution in the form $c(x)=\exp\left\{{i\over \epsilon}\,S(x)\right\}$.
Noting that $
\exp\left\{\epsilon \partial_x\right\}\,\exp\left\{{i\over
  \epsilon}\,S(x)\right\}=\left(1+i{\epsilon\over 2}S''\right)\exp\left\{i
S'\right\}\exp\left\{{i\over \epsilon}\,S(x)\right\}+{\cal O}(\epsilon^2),
$ where
$'\equiv\partial_x$, and expanding the ``action'' $S(x)$ as a power 
series in $\epsilon$, $S=S_0+i\epsilon S_1$,
we finally obtain
\begin{equation}\label{s0s1}
\epsilon^0:\quad \cos S_0'= {\lambda-\delta\,x\over 2F},\ \ \ {\rm and}\ \ \
\epsilon^1:\quad  S_1= -{1\over 2}\,\ln\left(F\,\sin S_0'\right).
\end{equation}
The ``classically allowed region'' corresponding to this problem 
is determined by the inequality 
\begin{equation}
\label{clas}x_-(\lambda)<x<x_+(\lambda),
\end{equation}
where  the turning points $x_{\pm}(\lambda)$
are solutions to
\begin{equation}
\label{eqx}\left(\lambda-\delta\,x \right)^2=\,4F^2(x)\, .
\end{equation}
\noindent
From \eqref{s0s1}, the semi-classical quantization 
rule for the spectrum can be shown
to be given by 
\begin{equation} \label{Genquant}
S(\lambda) = \int\limits_{x_{-}(\lambda)}^{x_{+}(\lambda)} \, dx \arccos
  \left(\frac{\lambda-\delta\,x}{ 2F(x)}\right) = 
\left(\alpha+\frac{1}{2}\right)\pi\, ,
\end{equation}
where $\alpha$ is a non-negative integer \cite{braun}. This relation 
yields the number of states with energy lower than $\lambda$, 
$\alpha(\lambda)$, whose derivative is the density of state given by
\begin{equation} \label{Gendens}
\rho(\lambda) = \left|\frac{d \alpha(\lambda)}{d\lambda}\right| = 
\frac{1}{\pi}
\int\limits_{x_{-}(\lambda)}^{x_{+}(\lambda)} \, \frac{dx}{v(x,\lambda)} 
\end{equation} 
where the ``velocity'' $v(x,\lambda)$ is given by 
$v(x,\lambda)=\sqrt{4F^2(x)-\left(\lambda-\delta\,x \right)^2}$. 
An approximate expression for the coefficients inside the allowed region is
\begin{equation} 
\label{coef}c(x,\lambda)={A\over\sqrt{v(x,\lambda)}}
\cos\left({1\over\epsilon}\,S_0(x,\lambda)+\theta\right)
\ \ \ {\rm where}\ \ \ S_0(x,\lambda)=\int\limits_{x_-(\lambda)}^x\,dx'\,
\arccos\left({\lambda-\delta\,x'\over 2 F(x')}\right).
\end{equation}
The angle $\theta$ depends on the
position of the turning points \cite{braun} but its expression is not needed
in what follows.  We will consider that, outside the allowed region,
$c(x,\lambda)$ decays fast enough to be neglected.  The normalization
constant $A$ is determined from the relation
\begin{equation} \label{norm}
1 = \sum_{p=0}^{P} |C_p(E)|^2 \simeq P\,
\int\limits_{x_{-}(\lambda)}^{x_{+}(\lambda)}
\frac{A^2}{v(x,\lambda)} \cos^2 \left({1\over \epsilon}
S_0(x,\lambda)+\theta\right) \, dx \simeq
P\,\frac{\pi}{2} A^2 \rho(\lambda).
\end{equation}
To obtain the last equality, we have replaced the rapidly
varying $\cos^2 \left(S_0/\epsilon\right)$ by its average value $1/2$.
This yields eventually
\begin{equation} \label{Cseminorm}
c(x,\lambda) = \left\{
\begin{array}{ccc}
\displaystyle \sqrt{\frac{2}{\pi P \rho(\lambda)v(x,\lambda)}} 
\cos \left({1\over \epsilon}S_0(x,\lambda)+\theta\right) & ,
& x_{-}(\lambda) \leq x 
\leq x_{+}(\lambda) \\
0 & , &\text{otherwise.}
\end{array} \right.
\end{equation}

\section{Atomic and molecular populations}

As already mentioned in the introduction, mean-field theory
predicts the existence of large-amplitude coherent oscillations between
atomic and molecular phases \cite{java}, while 
numerical simulations \cite{vardi} suggest that these oscillations are
strongly damped.  It is therefore important to clarify the
long-time behaviour of the atomic-molecular BEC. To do so, we
propose to evaluate the {\em dc-component} of the atomic population, 
$\overline{z}$.  
It follows from Eqs. \eqref{I00} and \eqref{Cseminorm}, that
this quantity can be expressed as
\begin{equation}\label{dc}
  \overline{z} = {1\over P} \,\sum_{\alpha}
  \left(C_{p_0}^{(\alpha)}\right)^2 
  \sum_{p} p
  \left(C_{p}^{(\alpha)}\right)^2\approx P^2 \!\!
  \int\limits_{\lambda_-(x_0)}^{\lambda_+(x_0)} \!\! d\lambda\, 
  \rho(\lambda)c^2(x_0,\lambda) \!\! 
\int\limits_{x_-(\lambda)}^{x_+(\lambda)}\!\! dx\,
  x\,c^2(x,\lambda)\approx {1\over\pi^2}
  \int\limits_{\lambda_-(x_0)}^{\lambda_+(x_0)}{1 \over
    v(x_0,\lambda)}\,
  {g(\lambda)\over\rho(\lambda)}\,d\lambda,
\end{equation}
where 
\begin{equation}
\label{g}g(\lambda)= \int\limits_{x_-(\lambda)}^{x_+(\lambda)}\,{x\,dx
  \over v(x,\lambda)} .
\end{equation} 
In deriving Eqs.\ (\ref{dc}) and (\ref{g}), we
have taken into account the fast oscillating nature of $C_p(E)$ as a 
function of $p$
for $E$ fixed (see Fig.\ref{fig1}), and therefore,
$\cos^2\left({1\over\epsilon}\,S_0\right)$ has been systematically
replaced by $1/2$.  Integrations in Eqs.\eqref{Gendens} and \eqref{g}
can be carried out (see e.g.\ \cite{grad}) and as a result, we
find the dc-component of the atom population to be
\begin{equation}\label{dcw}
\overline{z}={1\over\pi}\int\limits_{\lambda_-(x_0)}^{\lambda_+(x_0)}\,
\left\{x_t(\lambda)+\left(x_+(\lambda)-x_t(\lambda)\right){{\bf
    E}(m) \over{\bf K}(m)}\right\}\,{d\lambda \over v(x_0,\lambda)}
\end{equation} 
where $x_t(\lambda)$ is the third root of Eq.\eqref{eqx}
$\left(x_t\leq x_-\leq x_+\right)$, ${\bf K}(m) $ (${\bf E}(m)$) is
the complete elliptic integral of the first (second) kind 
\cite{abr} with parameter $m=(x_+-x_-)/(x_+-x_t)$,
and
$\lambda_\pm=\delta\,x_0\pm\,2F(x_0)$. For weak detuning,
$|\delta| < \sqrt{1-x_0}$, expression \eqref{dcw} is
well approximated by
\begin{equation}\label{dcwa}
\overline{z}\approx {2\over \pi}
  (1-\delta^2)\left\{{\arcsin
\left(\sqrt{{\lambda_+/(\lambda_+-\lambda_-)}}\right)\over
    \ln\left(8
      (1-\delta^2)/\lambda_+\right)}+
{\arcsin\left(\sqrt{{\lambda_-/(\lambda_--\lambda_+)}}\right)\over
    \ln\left(8 (1-\delta^2)/|\lambda_-|\right)}\right\}.
\end{equation}

Fig.\ \ref{fig2new} shows the dc-component of the expectation value of
the molecular mode population $z_{mol}\equiv \left(1 - \overline{z}
\right)/2$, in the resonant case $\delta=0$, as a function of the
initial value of the atomic mode population $z_0\equiv z(0)$. The
agreement between the numerical solution to Eq.\ \eqref{TTR} and the
analytical result \eqref{dcw} obtained within the discrete WKB
approach is quite remarkable.  We also present here the dc-component
of $\langle \hat{n}(t)\rangle$ obtained from the set of equations
(\ref{heq}) in the mean-field approximation.  In the resonant case
$\delta=0$ the atomic mode population is governed by the mean-field
equation
\begin{equation}
\label{mf}{d^2 z\over d\tau^2}+3 z^2-2z=0
\end{equation}
where $\tau=\chi\sqrt{\nu}\,t$ is a rescaled time. Under the initial
conditions 
$$
z(0)=z_0, \quad \frac{d z}{d \tau}{\Big |}_{\tau=0}=0
$$
the solution to Eq. (\ref{mf}) has the form
\begin{equation}\label{mfsol}
  z=z_0-z_0\,(1-q_+)\,\mbox{sn}^2\left(z_0\,\sqrt{z_0(1-q_-)}\,\tau\, 
{\Big|}\mu\right)
\end{equation}
where $\mbox{sn}(u|\mu)$ is the elliptic sine function with modulus
$\mu=(1-q_+)/(1-q_-)$ and where
$$q_{\pm}=\frac{1}{2\,z_0}\left(1-z_0\pm\sqrt{1+2 z_0-3
    z_0^2}\right).$$
It is straightforward to obtain from Eq.
(\ref{mfsol}) that the dc-component of the atomic population in the
mean-field approximation has the form
\begin{equation}\label{mfdc}
\bar{z}_{mf}=z_0\left(q_-+\,(1-q_-)\,\frac{{\bf E}(\mu)}{ {\bf K}(\mu)}\right).
\end{equation}
Fig. \ref{fig2new} shows that the numerical and analytical
WKB-solutions follow closely the mean-field solution (dashed line)
near the ends of the interval $z_0\in(0,1)$ ( for small and large
initial atomic populations). However, in the main part of the interval,
there is a significant discrepancy between numerical solution and
mean-field solution.  Note that the approximate formula for the
dc-component given in \eqref{dcwa} is also in a rather good agreement
with these results: the relative error does not exceed $0.08$.  Fig.
\ref{fig_mol3d} displays the dc-component of the molecular population
versus $\delta$ and $z_0$.  As observed from this graph, for a wide
range of initial atomic populations, $z_0$, a significant part of the
population is on average in the molecular mode.
  
\section{Population fluctuations}

To clarify the relevance of the
molecular population behaviour shown in Figs.\ \ref{fig2new} and
\ref{fig_mol3d} to other physical properties of the system, we need to
estimate the fluctuations of these quantities.  To this end we
consider the dc-component of the mean square fluctuation of the number
of atoms
\begin{eqnarray}\label{var}
\overline{\left(\Delta n\right)^2}\equiv
\overline{\langle\hat{n}^2(t)\rangle}-
\overline{\langle\hat{n}(t)\rangle}^2,\ \ \ 
\overline{\langle\hat{n}^2(t)\rangle}=\lim_{T \rightarrow \infty} 
\frac{1}{T} \int_{0}^{T} \,
\langle\hat{n}^2(t)\rangle\,dt.
\end{eqnarray}
From Eqs. (\ref{heq}) we find that the equation for the operator of
the number of atoms $\hat{n}(t)$ is
\begin{equation}\label{neq}{d^2\hat{n}(t)\over dt^2}+3\,
  \frac{\chi^2}{V}\,\hat{n}(t)^2+\left(\Delta^2-2\,
    \frac{\chi^2}{V}N\right)\,\hat{n}(t)
  - \frac{\chi^2}{V} N-2 \Delta\,H=0
\end{equation}
where $H$ is the Hamiltonian of the system (\ref{hamilt}).  
Now, averaging this last equation with
respect to any quantum state and taking its dc-part, yields the
following result {\em exact for any quantum state and any total number
  of particles $N$}
\begin{equation} \label{evolbar}
\overline{\langle\hat{n}(t)^2\rangle}/N^2=\frac{2}{3}\,(1-\delta^2)
\overline{z}+\frac{1}{3N} +\frac{2}{3}\,\frac{\Delta}{\chi^2 \,\nu}\,
\frac{H}{N} \,.  
\end{equation}  
From Eqs.\ (\ref{hamilt})
and (\ref{evolbar}), we find that, for an initial state given 
by Eq.\ (\ref{St1}) with $n_a=n_0$ ($n_b=(N-n_0)/2$), the mean
square fluctuation of the number of molecules in the atomic-molecular BEC
is
\begin{equation}\label{n2barvsnbar} 
\overline{\left(\Delta n_b\right)^2}/N^2 = \frac{1}{6}\, \left(1
-\delta^2\right)\,\overline{z}+\frac{1}{6}\,\delta^2\,z_0-
\frac{\overline{z}^2}{4}+\frac{1}{12 N}
\end{equation} 
with the dc-component of the atom population $\overline{z}$ given by
Eq.\ (\ref{dcw}). The relative fluctuation of the number of molecules
is given by the expression $v=\sqrt{\overline{\left(\Delta
      n_b\right)^2}}{\Big /}\overline{\left( n_b\right)}$.
From Fig.\ \ref{fig_var}, it can be seen that the fluctuations of the
number of molecules are of the same order of magnitude as their mean
values. It is worth recalling that the mean square fluctuations of the
number of particles in an ideal gas is of the order of $1/N$ and are
exceedingly small in macroscopic assemblies. On the other hand, the
mean square fluctuations of the populations of individual energy
states in an ideal Bose-Einstein gas are known to be greater than $1$
(see {\it e.g.}\ \cite{band}). The two-mode atom-molecule BEC 
is then an intermediate system with mean square
fluctuations both less than $1$ (see Fig. \ref{fig_var}) and 
greater than $1/N$. Given the large 
fluctuations experienced by the populations around their average values,
there is, strictly speaking, no convergence (in time) towards any state.

\section{Conclusion}
In summary, we have considered the two-mode model of atom-molecule
Bose-Einstein condensate, and studied quantum effects on atom-molecule
population oscillations. We used numerical simulations and analytical
approaches. We found numerically eigenvalues and eigenvectors of the
full N-body problem.  Analytical calculations were performed by using
the discrete WKB approach. We found excellent agreement between the
results of direct numerical simulations and analytical results. In
contrast, the results of the widely used mean-field approach
deviate significantly from the numerics for a wide range of atom
concentrations. We studied also quantum fluctuations in coherent
atom-molecule transformations. We found  that that the fluctuations
of the number of atoms (molecules) are of the same order of magnitude
as their mean values.  Thus, the two-mode BEC is seen to be in an
intermediate regime: molecules and atoms can still be treated as
separate phases, but their populations are highly fluctuating.

 \section*{Acknowledgements}
 Yu.B.G. thanks the Department of Physics, the Technical University of
 Denmark, for a Guest Professorship.  The authors acknowledge support by
 the EU LOCNET Project No. HPRN-CT-1999-00163.

\newpage

\begin{figure}
\includegraphics[scale=1]{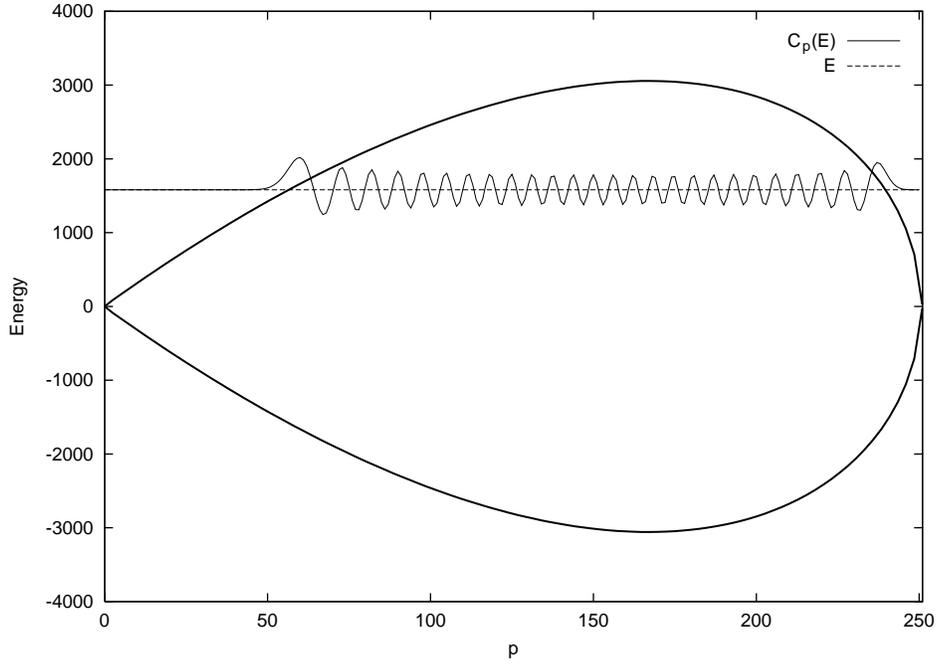}
\caption{\label{fig1} Coefficients $C_p(E)$ (arbitrary units) 
obtained from Eq. \eqref{TTR}
  for $\delta=0$ and $N=500$ ($P=250$) and placed at the corresponding energy
  $E\simeq 1580.8$ (thin dashed line). The allowed region is enclosed by 
  thick upper and lower lines. Coefficients $C_p(E)$ are
  clearly oscillating in this region and rapidly decaying outside.}
\end{figure}

\begin{figure}
\includegraphics[scale=1]{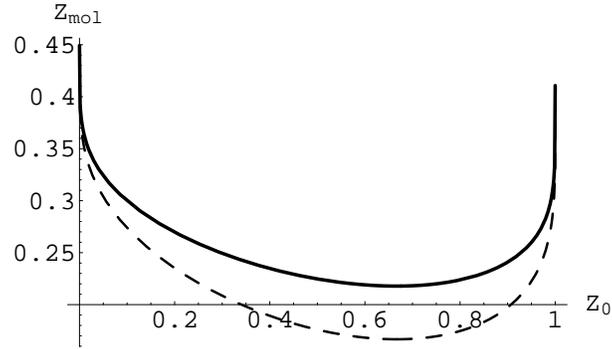}
\caption{\label{fig2new} Normalized dc-component of the molecular population 
 $z_{mol}=(1-\overline{z})/2$  versus the normalized initial number of atoms, $z(0)$ for a zero detuning parameter, $\delta=0$. Exact numerical results are 
obtained from Eqs. \eqref{TTR} and \eqref{I00} for $N=500$
cannot be distinguished from the analytical expression \eqref{dcw} shown as the solid line. The dashed curve represents the same quantity obtained in the framework of the mean field approach}
\end{figure}

\begin{figure}
\includegraphics[scale=1]{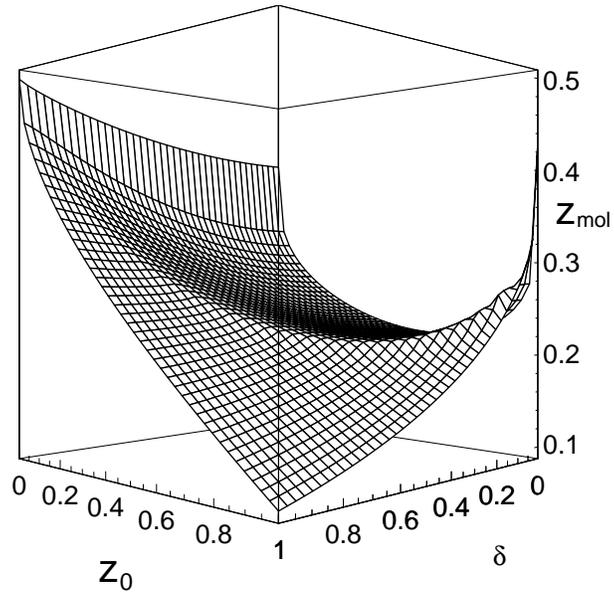}
\caption{\label{fig_mol3d} Dc-component of the molecular
  population, $z_{mol}=\left(1-\overline{z}\right)/2$, 
versus the detuning parameter $\delta$ and the initial value 
of the atomic population $z_0$.}
\end{figure}

\begin{figure}
\includegraphics[scale=1]{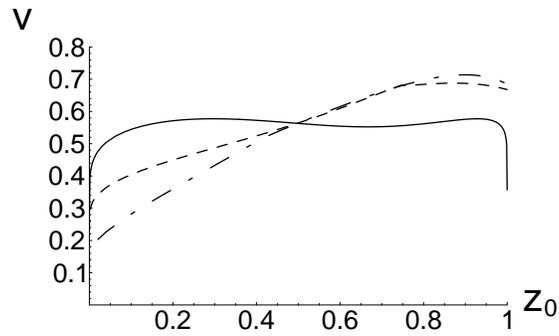}
\caption{\label{fig_var} Dc-component of the relative fluctuation
  of the number of molecules, $v=\sqrt{\overline{\left(\Delta
      n_b\right)^2}}{\Big /}\overline{\left( n_b\right)}$, 
  for different values of the detuning
  parameter $\delta$: $\delta=0$ (solid line), $\delta=0.5$ (dashed
  line) and $\delta=0.75$ (dot-dashed line) .}
\end{figure}  

\end{document}